\documentclass{emulateapj}

\shortauthors{Hinton et al.}

\begin{document}

\title{HESS J0632+057 : A NEW GAMMA-RAY BINARY?}

\author{
  J.A.~Hinton\altaffilmark{1},
  J.L.~Skilton\altaffilmark{1},
  S.~Funk\altaffilmark{2},
  J. Brucker\altaffilmark{3},
  F.~A. Aharonian\altaffilmark{4,5},
  G.~Dubus\altaffilmark{6},
  A.~Fiasson\altaffilmark{7},
  Y.~Gallant\altaffilmark{7},
  W.~Hofmann\altaffilmark{5},
  A.~Marcowith\altaffilmark{7},
  O.~Reimer\altaffilmark{2}
}
\altaffiltext{1}{School of Physics \& Astronomy, University of Leeds, Leeds LS2 9JT, UK}
\altaffiltext{2}{Kavli Institute for Particle Astrophysics and Cosmology, SLAC, 2575 Sand Hill
Road, Menlo-Park, CA-94025, USA}
\altaffiltext{3}{Universit\"at Erlangen-N\"urnberg, Physikalisches Institut, Erwin-Rommel-Str. 1,
D 91058 Erlangen, Germany}
\altaffiltext{4}{Dublin Institute for Advanced Studies, 5 Merrion Square, Dublin 2,
Ireland}
\altaffiltext{5}{Max-Planck-Institut f\"ur Kernphysik, P.O. Box 103980, D 69029 Heidelberg, Germany}
\altaffiltext{6}{Laboratoire d'Astrophysique de Grenoble, INSU/CNRS, Universit\'e Joseph Fourier, BP
53, F-38041 Grenoble Cedex 9, France}
\altaffiltext{7}{Laboratoire de Physique Th\'eorique et Astroparticules, CNRS/IN2P3,
Universit\'e Montpellier II, CC 70, Place Eug\`ene Bataillon, F-34095 Montpellier Cedex 5, France}

\begin{abstract}

The High Energy Stereoscopic System (HESS) survey of the Galactic plane has established the
existence of a substantial number ($\sim$40) of Galactic TeV
$\gamma$-ray sources, a large fraction of which remain
unidentified. HESS\,J0632+057 is one of a small fraction of these
objects which is point-like in nature ($<2'$ rms), and is one of only
two point-like sources that remain unidentified.  Follow-up
observations of this object with \emph{XMM-Newton} have revealed an
X-ray source coincident with the TeV source and with the massive star
MWC\,148, of the spectral type B0pe. This source exhibits a hard spectrum,
consistent with an absorbed power law with $\Gamma=1.26\pm0.04$, and
shows significant variability on hour timescales. We discuss this
spatial coincidence and the implied spectral energy distribution of
this object and argue that it is likely a new $\gamma$-ray binary
system with a close resemblance to the three known members of this
class and, in particular, to LS\,I\,+61\,303.  Further X-ray, radio and
optical observations of this system are needed to firmly establish
HESS\,J0632+057 as a new member of this rare class of Galactic
objects.

\end{abstract}

\keywords{binaries: gamma-ray}

\section{Introduction}

The emerging class of $\gamma$-ray binaries has three well-established
members, PSR\,B1259$-$63/SS\,2883 \citep{HESS:psrb1259} LS\,5039
\citep{HESS:ls5039p1,HESS:ls5039p2} and LS\,I\,+61\,303 \citep{
MAGIC:lsi61,MAGIC:lsi61b, VERITAS:lsi61}, all high-mass X-ray binary
systems \footnote{There is evidence of a single TeV \emph{flare} from
Cyg X-1, but $\gamma$-ray emission from this object is not firmly
established~\citep{MAGIC:cygx1}}.  All these objects show variable or
periodic TeV $\gamma$-ray emission with peak fluxes around
$10^{-11}$\,erg\,cm$^{-2}$\,s$^{-1}$ (1--10\,TeV). As such fluxes are
well above the sensitivity achieved by the HESS Galactic plane
survey \citep{HESS:scanpaper2}, the serendipitous discovery of new
$\gamma$-ray binaries might be expected in
HESS observations. Indeed, the TeV emission of LS\,5039 was
discovered in this way. Binaries are one of the few classes of sources
expected to appear \emph{point-like} for TeV instruments.  Of the
$\sim$40 unidentified TeV sources \citep{Jim:NJP}, only
HESS\,J1745$-$290 \citep{HESS:gcprl} (at the gravitational centre of
the Galaxy) and HESS\,J0632+057 \citep{HESS:0632} are unresolved. For
this reason alone HESS\,J0632+057 can be considered as a candidate for
a new binary system.

HESS\,J0632+057 was discovered in HESS observations of the region
of apparent interaction between the Monoceros Loop supernova remnant
and the star-forming regions of the Rosette Nebula. Three possible
associations were suggested by the HESS collaboration for this
object: the unidentified \emph{ROSAT} source 1RXS J063258.3+054857, the
unidentified \emph{EGRET} source 3EG\,0634+0521 and the massive star MWC\,148
(HD\,259440, spectral type B0pe).  To resolve the question of the
association of these three objects, and to probe the greater than 
3 keV emission
of the X-ray source, an \emph{XMM-Newton} observation of this region
was conducted in 2007, the results and implications of which are
described below.

\section{X-ray observations and analysis}

\begin{figure*}
\epsscale{1.11}
\plottwo{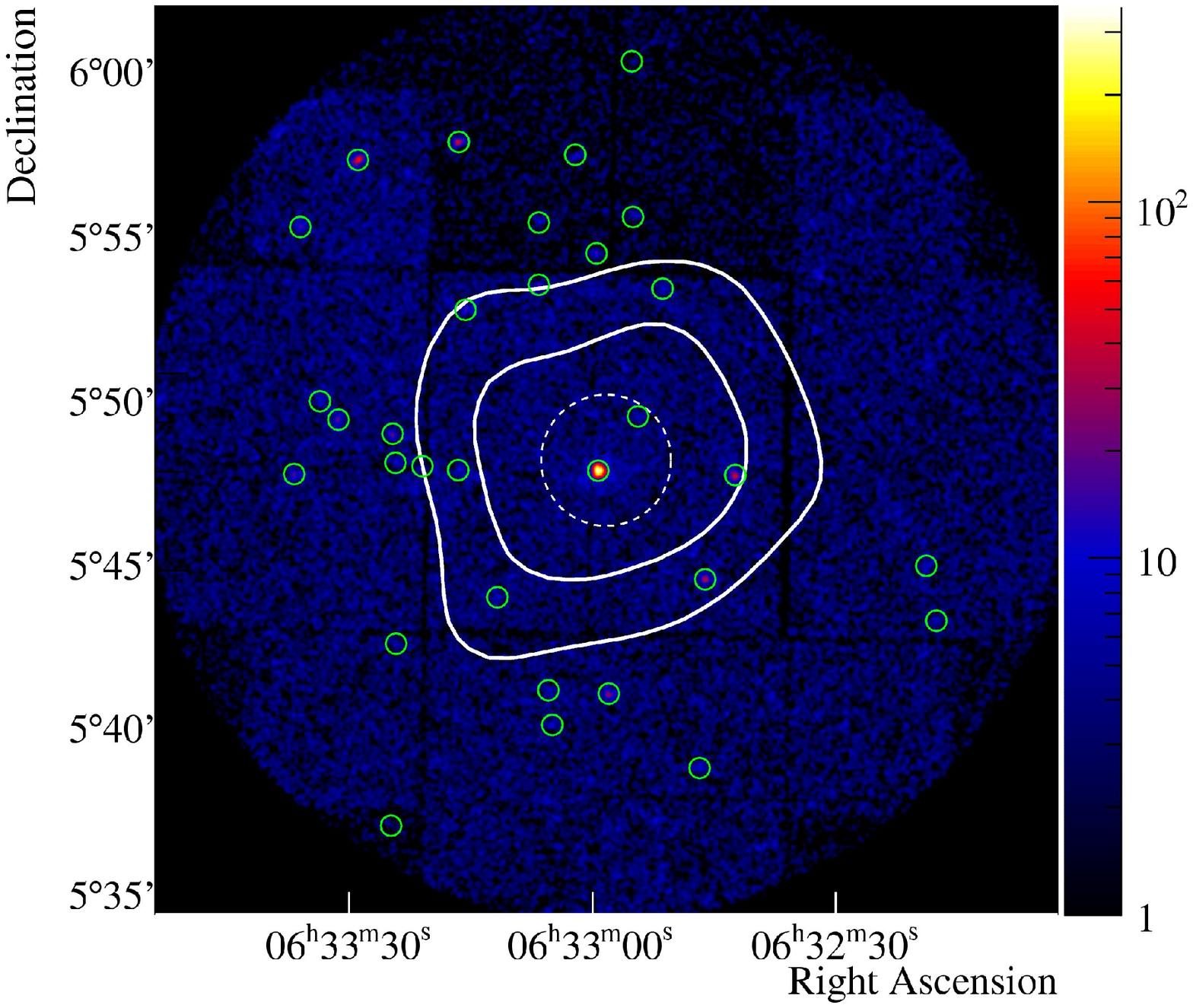}{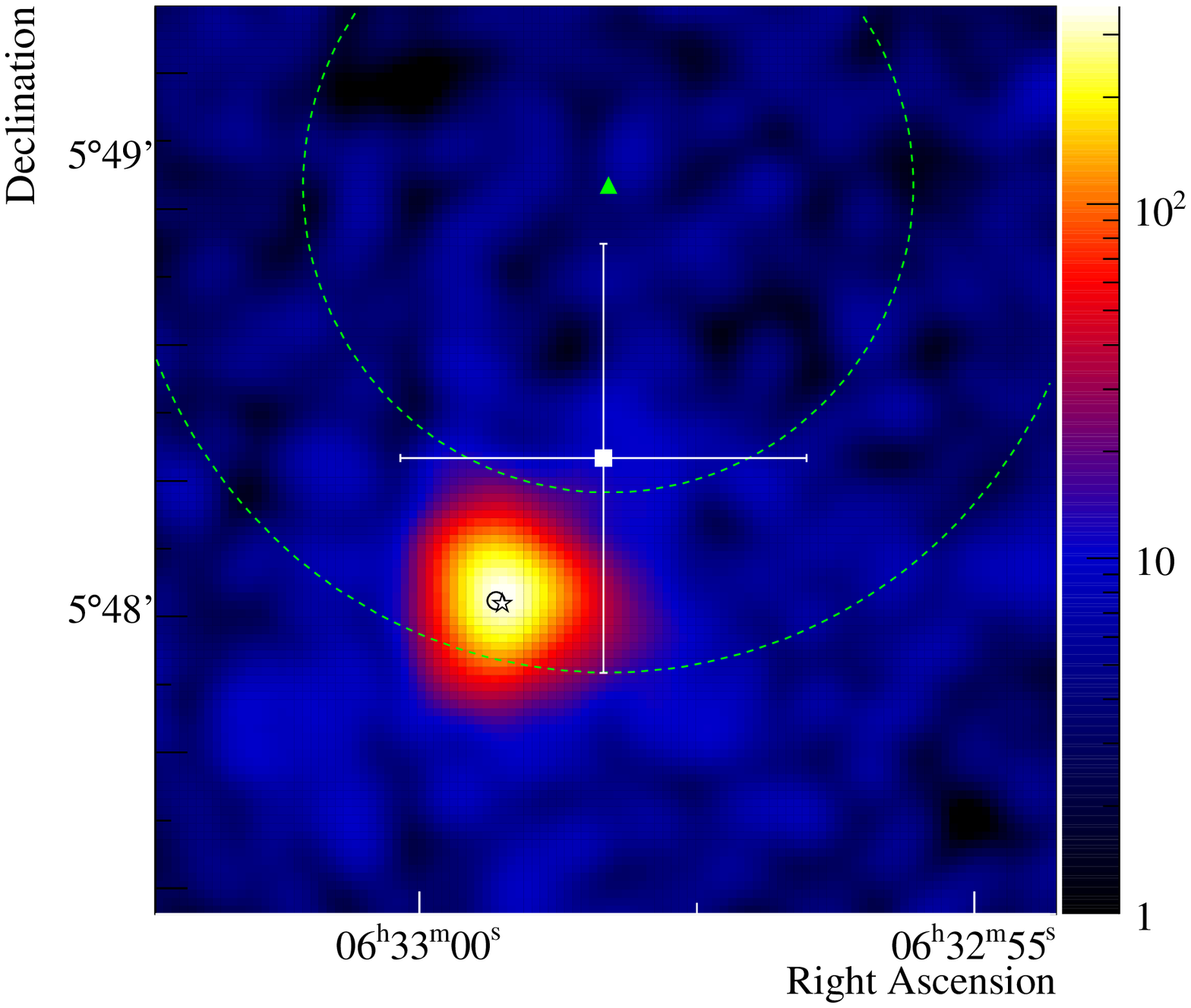}
\caption{\emph{XMM-Newton} combined MOS-1/MOS-2 image of the
  HESS\,J0632+057 region. \emph{Left:} Gaussian smoothed
  ($\sigma=3''$) count map (logarithmic colour scale) of the
  \emph{XMM-Newton} field of view, showing the 4 and 6 $\sigma$
  significance contours of the HESS source (solid lines,
  reproduced from \citet{HESS:0632}), the limit on the rms extent of
  the TeV source (dashed circle) and the $>$5$\sigma$ X-ray sources
  detected in this observation (green circles). \emph{Right:} The
  central $2'$ of the same map, showing the positional uncertainty on
  the centre-of-gravity of HESS\,J0632+057 (square marker with error
  bar), MWC\,148 (star), the best-fit position of
  XMMU\,J063259.3+054801 (open circle) and the ROSAT faint catalogue
  source 1RXS\,J063258.3+054857 (triangle and dashed circles for
  estimated 90\% and 99\% CL position errors).
\label{fig1}}
\end{figure*}

X-ray observations toward HESS\,J0632+057 were taken on 2007
September 17 with the EPIC camera of \emph{XMM-Newton} for
46~ks (observation ID\,0505200101). The analysis was carried out using the
\emph{XMM-Newton} Science Analysis Software (SAS, version 7.1.0). The
data were cleaned of a strong proton flare toward the end of the
observation, resulting in a useful exposure of 26~ks during which all
EPIC instruments were fully operational. The standard
\emph{edetect\_chain} tool was used to identify 31 $>$5$\sigma$
point-like sources in the field of view (FOV)(see Figure~\ref{fig1}).  
By far the brightest source is that found at 6$^{\rm h}$32$^{\rm
m}$59.29$^{\rm s}$~$\pm$~0.01$^{\rm s}$,
5$^{\circ}$48$'$1.5$''$~$\pm$~0.1$''$, within the HESS\,J0632+057
error box and offset by 0.6$''$ from the position of MWC\,148
\citep{Tycho2}, consistent with the nominal 1$''$ rms absolute
measurement accuracy of \emph{XMM-Newton}.  This object,
XMMU\,J063259.3+054801 (source \#1), exhibits an integrated signal of
2140~$\pm$~50 counts.  The source is unresolved and an upper limit on
the intrinsic rms width of 1.0$''$ (at 95\% confidence) has been
derived by convolving the simulated point-spread function with an
assumed Gaussian emission profile. Figure~\ref{fig1} illustrates the
spatial relationship of source \#1 to HESS\,J0632+057 and
1RXS\,J063258.3+054857.

\begin{figure}[h!]
\epsscale{1.07}
\plotone{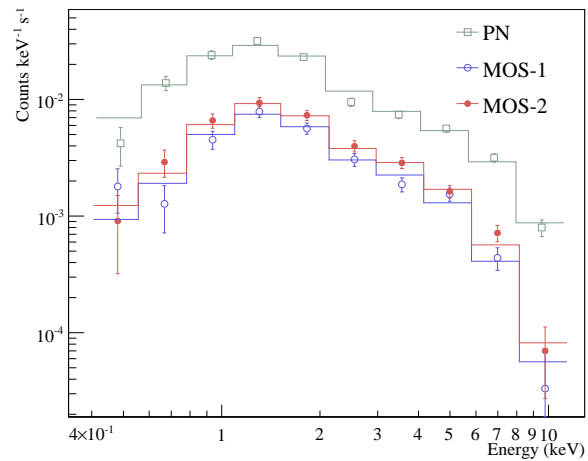}
\caption{Measured X-ray spectra for XMMU\,J063259.3+054801 from the 
three instruments of the EPIC camera. The histograms show the
best-fit forward folded absorbed power-law model. \label{fig2}}
\end{figure}

Energy spectra for source \#1 were extracted from a 25$''$ circular
region centered on the best-fit position for each of the EPIC CCD
cameras (MOS-1, MOS-2 and PN). For background estimation a 50--100$''$
annulus around the source was used for the MOS cameras and a 75$''$
circular region with an offset of 150$''$ from the source was
used for the PN (to avoid CCD chip edges). The resulting spectra were
simultaneously fitted with an absorbed power-law model, and are shown in
Figure.~\ref{fig2}. The best-fit model has a photon index
$\Gamma$\,=\,1.26$\pm$0.04, a column density $N_{\rm
H}$~=~(3.1$\pm$0.3)$\,\times\,10^{21}$~cm$^{-2}$, and 1~keV
normalization of (5.4$\pm$0.4)$\times$10$^{-5}$ keV$^{-1}$ cm$^{-2}$
s$^{-1}$, with $\chi$$^2$/dof~=~27.6/27. Whilst a single temperature
black body model is strongly excluded ($\chi$$^2$/dof~=225/27), two
temperature models provide a good fit with $kT_{1}\approx$0.4 keV and
$kT_{2}\approx$2 keV ($\chi$$^2$/dof~=15.8/25). The deabsorbed
1--10~keV flux from the power-law fit is $(5.3 \pm 0.4) \times
10^{-13}$ erg\,cm$^{-2}$ s$^{-1}$.  The count rate and hardness ratio
(HR2) of 1RXS\,J063258.3+054857 in the 0.4--2.4 keV \emph{ROSAT} band are
consistent (within large statistical errors) with the power-law model
for source \#1. There is, however, an apparently significant excess
over the model in the region below $0.3$ keV where \emph{XMM-Newton} has
limited sensitivity.  The catalogue position of 1RXS\,J063258.3+054857
lies 57$''$ from source \#1. Following \citet{Voges99} we estimate a
combined statistical and systematic error of 41$''$ (65$''$) at 90\%
(99\%) confidence. Given the consistency of the flux and spectrum in the
region of overlap, the marginal positional agreement, and in the
absence of any viable alternative counterpart, we tentatively identify
1RXS\,J063258.3+054857 with XMMU\,J063259.3+054801.

\begin{figure}[h]
\epsscale{1.08}
\plotone{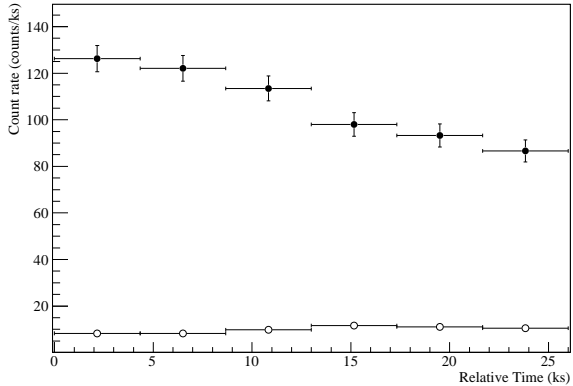}
\caption{X-ray light-curve for XMMU\,J063259.3+054801 for the combined
EPIC cameras. Background-subtracted signal (closed circles) and
estimated background (open circles) are shown.\label{figLC}}
\end{figure}

A search for variability of source \#1 has been conducted, again using
all three EPIC cameras. Figure~\ref{figLC} shows combined EPIC
light curves for the signal and background regions described earlier.
We find evidence of variability on a timescale comparable with the
observation: a fit to a constant value yields a chance probability of
$2\times10^{-9}$. The data are consistent with a linear fit which
implies a decrease of $39\,\pm\,5$\% in flux over the 26~ks
observation.  We find no evidence of a change in spectral shape when
splitting the data into two 13~ks slices. With $N_{\rm H}$ fixed to
$3.1\,\times\,10^{21}$~cm$^{-2}$ the spectral indices of the first and
second segments are $1.25\pm0.04$ and $1.28\pm0.06$ respectively. With
both $N_{\rm H}$ and the spectral index fixed to the full-dataset values
the best-fit normalization falls from (6.1$\pm$0.2)$\times$10$^{-5}$
keV$^{-1}$ cm$^{-2}$ s$^{-1}$ to (4.6$\pm$0.2)$\times$10$^{-5}$
keV$^{-1}$ cm$^{-2}$ s$^{-1}$ in the second slice.

\section{Discussion}

A fundamental question in the interpretation of XMMU\,J063259.3+054801
is the physical process responsible for the X-ray emission. Given the
existence of TeV emission from this region, it seems natural to
attribute the power-law X-ray emission as synchrotron emission from
ultrarelativistic electrons. However, spectra resembling that
observed for source \#1 (consistent with two temperature blackbody
models with $kT_{1}\sim$0.5~keV and $kT_{1}\sim$2~keV) have been
observed for isolated magnetic Bp stars, such as $\sigma$\,Ori\,E, and
been attributed to shock heating of magnetically confined winds
\citep[see][and references therein]{Townsend}. It seems possible that
MWC\,148 is a system of this type, and that the observed X-ray
emission is thermal in origin. However, it is difficult to explain the
TeV $\gamma$-ray emission in such a scenario. In the discussions that
follow we assume that the observed X-ray emission is nonthermal in
origin. If this is not the case, then the observed flux provides an
upper limit on the nonthermal X-ray flux of HESS\,J0632+057.

The probabilities of chance associations of HESS\,J0632+057 with
MWC\,148 and 1RXS J063258.3+054857 have been estimated as $10^{-4}$
and $10^{-3}$ \citep{HESS:0632} respectively. The chance coincidence
of a massive star with the 1$''$ error box of the brightest (and
the hardest spectrum) source in an \emph{XMM-Newton} observation is even
less probable ($\sim10^{-6}$). The least secure association is that
with 3EG\,0634+0521, as $\sim$10\% of all Galactic TeV sources lie
within the 99\% confidence contour of a 3EG source
\citep{GeVTeV}. Furthermore, the binary system SAX\,J0635+0533
(composed of a 34~ms period pulsar and a Be star) has been suggested
to power most or all of the flux of 3EG\,0634+0521~\citep{SAX0635}.
However, as 3EG\,0634+0521 is flagged as extended and/or confused it
seems plausible that HESS\,J0632+057 may contribute to this
emission. The relative contributions of these and other sources to
3EG\,0634+0521 should be established by the first months of data from
\emph{GLAST}. We note that there is no evidence of TeV emission from
SAX\,J0635+0533 \citep{HESS:0632}.

If all the apparent associations of HESS\,J0632+057 are correct then
its spectral energy distribution (SED) bears a close resemblance to
that of the known TeV binaries. All have hard ($\Gamma<2$) X-ray
spectra and significantly softer TeV spectra.  Furthermore, X-ray
variability with timescales comparable to those observed here for
source \#1 has been observed in LS\,I\,+61\,303
\citep{Esposito}. Associations with GeV $\gamma$-ray sources also
exist in the cases of LS\,5039 (3EG\,J1824$-$1514) and LS\,I\,+61\,303
(3EG\,J0241+6103).  These systems are powered by either a relativistic
pulsar wind or a wind-accretion powered jet \citep[see
e.g.][]{BinariesReview}.  An assessment of the possible power source
of HESS\,0632+057 requires an estimate of the distance to the system,
Using the apparent visual magnitude $m_{V}=9.16$ of MWC\,148
\citep{Tycho2}, assuming an absolute magnitude $M_{V}=-4.0$, and
accounting for reddening assuming an intrinsic $M_{B}-M_{V}$ of -0.3,
we find a distance of $\approx$1.5~kpc, consistent with that of the
Rosette Nebula \citep{RosetteDist}
\footnote{Following \citet{ryter}, the implied optical extinction of
$E_{B-V} = 0.75$, is compatible with the best fit $N_{H}$ for source
\#1, supporting the physical association of the two sources.}.  At
this distance, the $1''$ limit on the intrinsic size (and the similar
limit on the projected displacement from MWC\,148) of the X-ray source
implies an origin of the emission within $\approx2\times10^{16}$~cm
of the star and hence a radiation density of greater than 10$^{6}$~eV~cm$^{-3}$
\footnote{Causality arguments limit the size of the X-ray emission
region (rather than the distance from the star) to $r_{X}<c\Delta t$,
where $\Delta t$ is the variability timescale of $\sim$30~ks, leading
to $r_{X}<10^{15}$ cm.}.  In the presence of such a strong, high-energy 
($kT$\,$\sim$\,3~eV) photon field, inverse Compton (IC)
scattering of $\sim$TeV electrons must occur in the deep Klein--Nishina
(KN) regime. IC cooling is dominant if $f_{\rm KN}(E_{e})U_{\rm rad} >
U_{\rm mag}$, where $f_{\rm KN}(E_{e})$ is the suppression factor due
to the KN effect. Following \citet{Moderski}, $f_{\rm KN}(E_{e})
\approx (1.0 + (4E_{e}\times2.8kT)/(m_{e}c^{2})^{2})^{-1.5}$, and for
$kT=3$~eV and $E_{e}\,=\,1$~TeV, $f_{\rm KN}\sim10^{-3}$.  If IC
cooling in the KN regime is dominant, then a hard X-ray spectrum and
softer TeV $\gamma$-ray spectrum, as observed in this case, can be
expected. Such a scenario has been recently discussed for a pulsar
wind nebula in the central parsec of our galaxy~\citep{GCPWN} and for
binary systems such as PSR\,B1259$-$63/SS\,2883~\citep{Khangulyan07}.

Assuming for the moment a distance of the emission region from the
star of a few AU, radiation densities of $U_{\rm rad} \sim$1
erg~cm$^{-3}$ are expected. The relative flux of X-rays and TeV
$\gamma$-rays then implies $B\sim 5
(f_{KN}\,F_{X}/F_{\gamma})^{0.5}\,\mathrm{G}\,\sim\,70$ mG.  For these
radiation and magnetic fields 1 TeV electrons produce X-rays of a few
keV and $\gamma$-rays of almost TeV energy (if the IC scattering
proceeds in the deep KN regime). The cooling time
($E/\mbox{d}E/\mbox{d}t$) for these electrons is then $\sim$20~ks,
compatible with the X-ray variability observed for source \#1.
Furthermore, in contrast to synchrotron-dominated cooling, the cooling
timescale is almost constant with energy in the region probed by
synchrotron X-rays and no spectral variability is expected, again
compatible with this observation. We note that in this scenario
correlated X-ray and TeV $\gamma$-ray variability is expected.

\begin{figure*}[t]
\epsscale{1.04}
\plotone{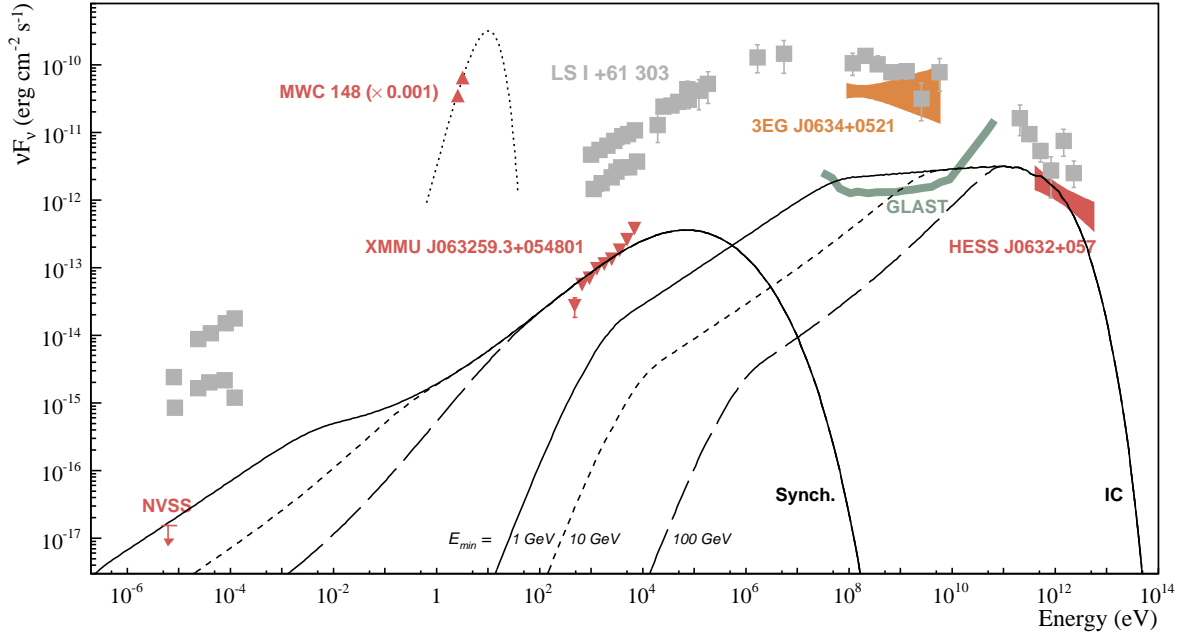}
\caption{Spectral energy distribution of HESS\,J0632+057 / XMMU\,J063259.3+054801
(red symbols/regions) compared to that of the $\gamma$-ray binary system LS\,I\,+61\,303 
\citep[grey symbols/regions, see][and references therein]{Chernyakova} and the estimated 
1 year GLAST sensitivity for this region (thick solid line). The solid, short-dash 
and long-dash curves show the synchrotron and IC components of a simple time-dependent 
one-zone model with $E_{\rm min}$ set to 1 GeV, 10 GeV and 100 GeV, respectively 
- see text for details.\label{fig_sed}}
\end{figure*}

An illustrative one-zone model with these parameters ($B=70$~mG,
$U_{rad}=1$ erg~cm$^{-3}$) is shown in Figure~\ref{fig_sed}. Particles
are injected with a spectrum $dN/dE \propto E^{-\alpha}
\exp{(-E/E_{\rm max})}$, with $\alpha=2$ and $E_{\rm max} = 10$ TeV,
and cool continuously. The SED reaches an equilibrium state (as shown)
for timescales $\gg 20$ ks. Lower values of $E_{\rm max}$ are
difficult to reconcile with the highest-energy X-ray and $\gamma$-ray
data points. The effect of changing the minimum energy of injected
electrons is shown in Figure~\ref{fig_sed}. We note that values as high
as 100~GeV (long-dashed curve) might be expected in the binary PWN
scenario: \emph{GLAST} is well suited to probe this region (see sensitivity
curve in Figure~\ref{fig_sed}).  If genuinely associated, the flux level
of 3EG\,0634+0521 may represent a higher state of this variable
object, or indicate a softer electron spectrum in the GeV range.

The power requirements of the observed emission place limitations on
the possible energy source in the system. The model shown in
Fig.~\ref{fig_sed} requires a power of $10^{34}$ erg s$^{-1}$. The
luminosities in individual wavebands are $\approx1.3\times10^{32}$
erg s$^{-1}$ (1--10 keV), $\approx$3$\times10^{34}$ erg s$^{-1}$
(0.2--2~GeV), and $\approx$5$\times10^{32}$ erg s$^{-1}$
(0.4--4~TeV). These luminosities should be compared to the available
kinetic power of the Be-star wind \citep[$\cal O$($10^{34}$) erg
s$^{-1}$, see e.g.][]{Waters}, the wind accretion luminosity on to an
(unseen) compact companion \citep[$\sim$$10^{35}$ erg s$^{-1}$, see
][]{AccretionBook}, and the spin-down power of an (unseen) young pulsar
companion \citep[$\dot{E}\approx$$8\times10^{35}$ erg s$^{-1}$ e.g. in PSR\,B1259$-$63, ][]{Johnston}.

\section{Summary and Outlook}

It seems very likely that the sources identified as HESS\,J0632+057
and XMMU J063259.3+054801 originate in the same astrophysical object
and that this object is associated with the Be-star MWC\,148. Whilst
the nature of this object remains uncertain, it is unlikely that an
isolated star can provide the necessary power or accelerate particles
to the required maximum energy of $\gg$1~TeV. MWC\,148 is, therefore,
likely part of a binary system. By analogy with known $\gamma$-ray
binaries, which exhibit similar X-ray and TeV spectra, the unseen
companion of MWC\,148 is likely a young pulsar or an accreting black
hole or a neutron star driving a jet.  The possible association of this
object with 3EG\,0634+0521 will be tested in the near future using
\emph{GLAST}, which should achieve a source location accuracy of
$\approx$100$''$ after 1 year.  The observed emission and variability
appear consistent with synchrotron and IC emission from a population
of relativistic electrons in a region within a few AU of the star.
Continued radio--$\gamma$-ray monitoring of this object is proposed
to establish the existence of periodicity and/or correlated
variability and to confirm or refute the binary nature of this object.

\acknowledgments The authors would like to thank Willem-Jan de Wit, 
Teddy Cheung and J\"orn Wilms for useful discussions. JAH is supported by a UK 
Science and Technology Facilities Council (STFC) Advanced Fellowship.

{\it Facilities:} \facility{XMM-Newton}

\end{document}